\newif\ifREFEREE \REFEREEfalse

\ifREFEREE 
   \documentstyle[epsfig,referee]{l-aa}
   
\else
   \documentstyle[epsfig]{l-aa}
   
\fi

\def \( {\left(}
\def \) {\right)}
\def \Rv {R_{V}}
\def \Av {A_{V}}
\def \mic {\ \mu \mbox{m}}

\begin{document}

\thesaurus{09  (09.03.1 ; 09.04.1 ; 09.09.1 )}

\title{Extinction map of Chamaeleon I molecular cloud with DENIS star counts\thanks{Based on observations collected at the European Southern Observatory, La Silla, Chile}} 

\author{L. Cambr\'esy\inst{1} \and N. Epchtein\inst{1} \and E.Copet\inst{1} \and B. de Batz\inst{1} \and S. Kimeswenger\inst{2} \and T. Le Bertre\inst{1} \and D. Rouan\inst{1} \and D. Tiph\`ene\inst{1}}
          
\institute{Observatoire de Paris, CNRS URA 264 and URA 335, F-92195 Meudon Cedex, France 
	  \and Institut f\"ur Astronomie der Leopold--Franzens--Universit\"at Innsbruck, Technikerstra{\ss}e 25, A--6020 Innsbruck, Austria}

\offprints{Laurent Cambresy, Cambresy@denisezg.obspm.fr}
            
\date{Received ....}
\maketitle

\begin{abstract}

Massive, large scale star counts in the $J$ (1.25 $\mic$) band provided by the
Deep Near Infrared Survey of
the Southern Sky (DENIS) are used for the  first time to draw out an
extinction  map of the Chamaeleon I dark cloud. 
We derived a $2'$ resolution  map of the cloud from $J$ star counts within
an area of $1\fdg5 \times 3\degr$ around the centre of the cloud using 
an adaptive grid method and applying a wavelet decomposition. Possible
contaminating  young stellar objects within the cloud are removed, although
they are shown to have a negligible effect on  the counts. A comparison of our
extinction map with  the  {\it cold} contribution of the  IRAS 
$100 \mic$
emission  shows an almost perfect matching. It is shown that $J$ star counts
supersede optical counts on Schmidt plate where $\Av > 4$.


\end{abstract}

\section{Introduction}

The Chamaeleon I cloud is the most obscured  region of the Chamaeleon 
dust-molecular complex. An  extinction map of this cloud has been drawn  
out  by Gregorio Hetem et al. (1988) using star counts on ESO(B) plates.
It shows  a maximum  of $\Av \sim 6$, and the extinction profile  across
the cloud centre shows an abrupt growth between 2 and 6 $\Av$, followed by 
a plateau at the  peak of extinction which looks like a saturation.
As  stars become  too scarce, optical  counts can no longer be used
to derive  the extinction accurately. Near infrared (1-2$\mic$) star counts
are more appropriate to probe regions where $A_V \ga 4$  since  an extinction of
10 visual magnitudes drops to only $\approx$ 3 magnitudes in the $J$ band at 
1.25 $\mic$. 
Massive star counts in the near infrared  are  made possible for the first time thanks to 
the large scale near infrared survey, DENIS,  currently in progress (Epchtein, 
1997). The aim of this paper is to investigate in detail the extinction
toward the Cham I cloud using this new wealth of data.

The Cham I dark cloud is located at $b = -16\degr$ and
its current distance estimate is 140 pc (Whittet et al, 1987).
Its high galactic latitude implies a small number density of background stars which
limits the spatial resolution of the extinction estimation, but, on the other
hand, the probability of crossing several clouds on the line of sight is low.

\section{Observations}

The observations presented here have been collected as part of the DENIS survey 
between January and May 1996 at La Silla (Chile) using 
the ESO 1 meter telescope equipped with the specially designed 3-channel  camera
(Copet, 1996).  They cover an area of $1\fdg45 \times 2\fdg94$ centered
at $\alpha=11^{\rm h}06^{\rm m} \, ,\, \delta=-77\degr 30^{\rm m}$
(J2000) in three bands, namely $I$ (0.8$\mic$), $J$ (1.25$\mic$) and
$K_{\rm s}$ (2.15$\mic$).
They  consist of 13 strips each involving 180 images of $12'\times12'$ taken
at constant RA  along an arc of $30\degr$ in declination. The overlap 
between two  adjacent strips reaches $75 \%$ in the Cham I cloud because of the 
proximity of the south pole. Wherever  a star is  picked up in two adjacent
strips, position and flux values are averaged. Limiting magnitudes
are 18, 16 and 14 at $3\sigma$ in $I$, $J$ and $K_{\rm s}$ bands, respectively, and
position accuracy is $1 \arcsec$ in both directions.

\section{Star count method}

Usually, the extinction is evaluated by  comparison of star counts in  the 
absorbed region and a nearby area assumed to be free of obscuration (Wolf
diagram method). Star counts are performed by adding up the stars
up to a given magnitude (or in a given magnitude range, e.g., $\pm \frac{1}{2}$)
within a grid of fixed squares. The step of the grid is a compromise between
the stellar density and the spatial resolution. In other words, the spatial
resolution is underestimated wherever the extinction is low, while in highly
obscured areas, the content of several cells must be merged, in order to pick
up enough stars.

We have developed a new method to investigate the extinction across a cloud
which consists in replacing usual star counts by an estimation of the local
projected star density obtained by measuring the mean distance of the $x$
nearest stars. The most important advantage of this method is to match the
local extinction: it corresponds to a star count with adaptable square size. 
Another very interesting advantage of the method is to provide
a map with white noise. Therefore, we can simply estimate the noise by computing
the standard deviation $\sigma$ of the mean distance on a part of our map
with no signal. 

We obtain a map where each point represents the square root of the local
density. The extinction is then easily derived by the relation :
\begin{eqnarray}
A_{\lambda}=\frac{1}{a} \log \( \frac{\overline{d_{\rm {cl}}}}{\overline{d_{\rm cp}}}\) ^{2} 
\end{eqnarray}
where $a$ is defined by :
\begin{eqnarray}
\log \( \overline{d_{\rm cp}}\) ^{-2}=a \times m_{\lambda} + b
\end{eqnarray}

\noindent
where $m_{\lambda}$ is the magnitude,
$\overline{d_{\rm cl}}$ the mean distance of the $x$ nearest stars in the cloud and
$\overline{d_{\rm cp}}$ the mean distance of the $x$ nearest stars in a comparison
field supposed unobscured. We found $a=0.34$.
We verified that the relation (2) is correct up to $J=16$, i.e. our limit of completeness.
Actually, we do not need a reference field to derive the extinction because
the available data cover $30\degr$ of latitude and, thus, widely exceed the
cloud dimension. 
So, we plotted the density versus the galactic latitude to interpolate the density
inside the cloud for no extinction. 
Then we convert $A_{J}$ into visual extinction using the extinction law of
Cardelli et al. (1989) for $\Rv= \frac{\Av}{E_{B-V}}=3.1$. 
So $\frac{A_{J}}{\Av}=0.282$.

\ifREFEREE
   \begin{figure}[p]
\else
   \begin{figure}[htb]
   \epsfig{figure=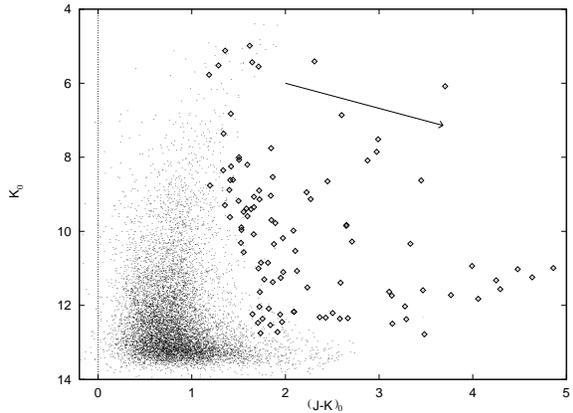,height=5.5cm} 
\fi
\caption[]{Colour--magnitude diagram of dereddened stars in an area of 4.3 deg$^2$. Diamonds represents young stellar object candidates (100 on 30\,000 stars), the arrow corresponds to the extinction vector of $\Av =10$}
\label{col-mag}
\end{figure}

\ifREFEREE
   \begin{figure}[p]
\else
   \begin{figure}[htb]
   \epsfig{figure=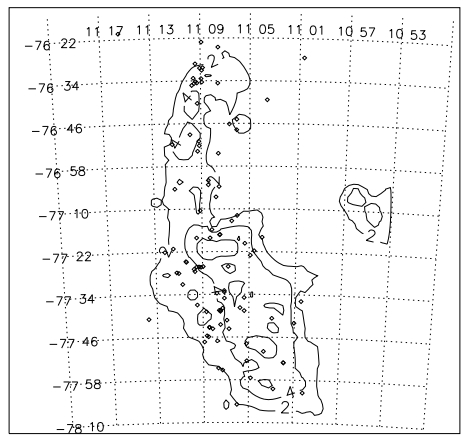,height=6.0cm} 
\fi
\caption[]{Spatial distribution of young stellar object candidates. Extinction isocontours at 2, 4 and 7 $\Av$ are overplotted}
\label{flag}
\end{figure}

A limitation to star counts behind molecular clouds, is the possible presence
of young stars embedded inside the cloud itself. To draw out a reliable
extinction estimate, the counts must be dominated by background stars.
Therefore, we have attempted to remove these {\em spurious}, although
interesting, objects using a colour excess criterion. A first iteration of
the extinction estimation is carried out without taking into account these
objects. Then, this map is used to deredden all stars, individually. Their
colours $J-K$ are compared to the main sequence star at the cloud distance
in a colour--magnitude diagram ($K$ versus $J-K$). Then we flag each star 
which would have an extinction even greater than 5 $\Av$ and which are
likely to be  intrinsically very red objets. The colour--magnitude
diagram is presented in Fig.~\ref{col-mag}. The spatial distribution of these
flagged stars (Fig.~\ref{flag}) show essentially two clusters
near the two $A0$ stars HD\,97300 (at the north) and HD\,97048 (in the centre). 
This result is in agreement with several studies of T--Tauri associations
in the Cham I with IRAS (Assendorp et al., 1990) and ROSAT (Lawson et al., 1996). 
We find also several red stars farther away from
the dense cores of the cloud. They could be real objects (T--Tauri, red giant),
or their position in the diagram might be unreliable because of the uncertainty on the
$J-K$ colour. We emphasize the fact that our criterion identifies only the
classical T--Tauri, but not the so called {\it weak--line} T--Tauri which have
no or little infrared excess. Nevertheless, this type of T--Tauri does not
concentrate in the dense cores as the classical one does, so they have probably
little effect on star counts. Finally, this operation removes only $0.5 \%$
of stars ($\approx 100 \mbox{ over } 30\,000$). Counts are, thus, strongly
dominated by background stars. Then, we have built an extinction map from the
cleaned counts.

We can consider our map as a digitized {\it image} which allows to use current
technics of image processing such as the wavelet transform to restore the image
and to filter the noise (Starck \& Murtagh, 1994). We  apply the {\it \`a trous}
wavelet transform algorithm to split-off the image into 4 wavelet planes.
The decomposition is made by  convolving the image  by a low--pass filtering
matrix. The difference between the  original image and the result of the first
convolution gives the first plane of the wavelet transform which  corresponds
to the high frequency plane. Further iterations of this process provide the 4
wavelet planes and the final smooth plane.

Thus, we can use the high frequency plane to identify aberrant points and 
remove them in the final image in order to eliminate their contribution in all
the planes, by replacing the bad pixels by the average of the surrounding 8
pixels. We are conscious that this process might result in a loss of
information, but less than $0.05 \%$ pixels are actually corrected in this way.

Lastly, we filter each wavelet plane using the following method. 
The noise on star counts is poissonian,
but taking the logarithm, as defined in equation (1) 
changes the statistical properties  which are no longer poissonian. 
A Poisson noise having the standard deviation  estimated in a region of the map
with no signal is simulated. Then we take its logarithm 
and we decompose this simulated noise into wavelet planes.
The estimation of the standard deviations $\sigma_{i}$ on each plane allows
an adaptable thresholding. Then we filter each plane at $3 \sigma_{i}$.

\section{Results}

In order to optimize the resolution, we choose the $J$ band and a number 
$x=20$ stars to estimate the local density. The extinction is characterized
at each point by the mean distance to the 20 nearest stars. 
In the $J$ band the resulting spatial resolution is $1'$ for low extinction 
($\Av \sim 1$) and $2'30''$ for $\Av \sim 8$. In the I band we find more
than $4'$ and the $K_{\rm s}$ band is not sensitive enough. 
The final result is the extinction map presented in 
Fig. \ref{map} with the IRAS 100$\mic$ emission isocontours in which a
{\it warm} contribution has been subtracted (see below). This map in false
colours results from the recombination of the 4 wavelet planes.

\ifREFEREE
   \begin{figure}[p]
\else
   \begin{figure}[htb]
   \epsfig{figure=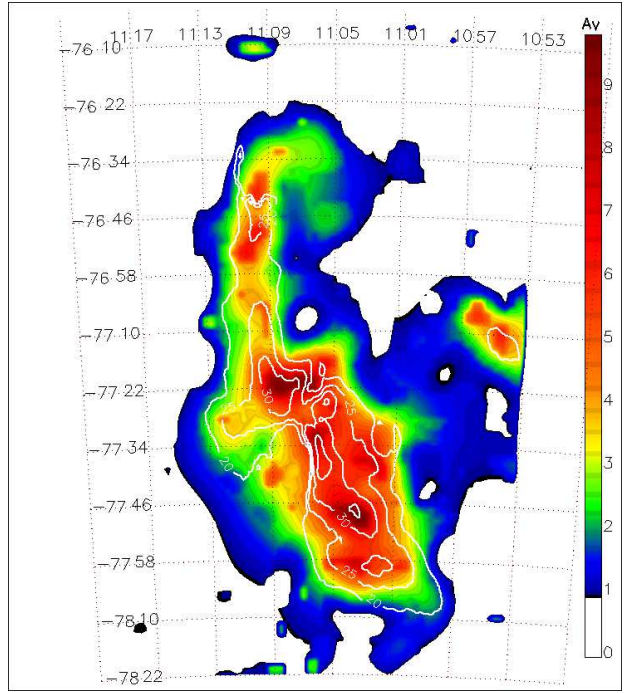,height=9.2cm}
\fi
\caption[]{Extinction map derived from $J$ band stellar counts and cold IRAS 100$\mic$ emission isocontours at 20, 25, 30, 35 MJy.sr$^{-1}$ (see text)}
\label{map}
\end{figure}

The standard deviation of the map is $\sigma=1.4$ magnitude. The resolution
for high extinction allows to clearly identify 4 distinct maxima greater
than 7 which were not resolved by the Schmidt plate analysis.
The maximum is $\Av \sim 9.8$. There is no evidence of saturation. Using the
relation between visual extinction 
and N$_{\rm H}$ column density (Savage and Mathis, 1979) we derive the 
mass $M$ of the cloud for $\Av > 2$ :

$
M=\alpha \mu d^2 \, \frac{{\rm N}_{{\rm H}_{\rm I}}+2N_{{\rm H}_2}}{\Av} \sum_i \Av^i=280 M_{\sun}
$

\noindent
where $\alpha$ is the angular size in square radian, $d$ the distance to the
cloud, and $\mu$ the mean particle mass. To calculate $\mu$ we
consider a gas composed of $91.5 \%$ of hydrogen (H$_{2}$+HI)
and $8.5 \%$ of helium. The mass cannot be estimated in a straightforward way
for $\Av<2$ because the slow variation of the extinction and the lower 
sensitivity in  the $J$ band induce important consequences on the value of the
solid angle $\alpha$ which delineates the absorbing region, but the larger 
uncertainty on the mass determination comes from the distance  estimate.

The extinction map that we have drawn out and the 100$\mic$  map of IRAS
have  a comparable spatial resolution. It is therefore tempting to 
cross-correlate the 2 maps to derive some properties of the dust grains.   
According to Laureijs et al. (1991) 
the IRAS 100$\mic$ flux consists of a {\it cold} and a {\it warm}
components which can be  split off into two components using the 60/100$\mic$
colour temperature. 
Following Boulanger et al. (1997), the cold 
contribution can be written:
$$
I_{\rm cold}(100\mic)=1.67 \times ( I_{\nu}(100\mic) - 5 \times I_{\nu}(60\mic))
$$
Figure \ref{map} shows that our extinction map and the cold 100$\mic$
emission are in very good agreement.
Note that the two areas with no IRAS contour ($11^{\rm h}10^{\rm m},\, -76\degr 
34^{\rm m} \mbox{ and } 11^{\rm h}08^{\rm m},\, -77\degr34^{\rm m}$)
are due to the presence of peculiar objects such as the {\it Infrared
Nebula} (Schwartz \& Henize, 1983) and very young stellar objects with outflows (Jones et 
al., 1985). This excellent correlation suggests that the $J$ extinction and
the cold 100$\mic$ emission have the same origin, a result in agreement with the
D\'esert at al. (1990) dust model which shows that the 100$\mic$ emission and
the near infrared extinction are both caused by big grains. The warm
component contribute strongly to $100\mic$ emission, but not much to the
extinction. A plot of the cold IRAS $100\mic$ emission versus visual 
extinction is presented in Fig. \ref{iras_av}.

\ifREFEREE
   \begin{figure}[p]
\else
   \begin{figure}[htb]
   \epsfig{figure=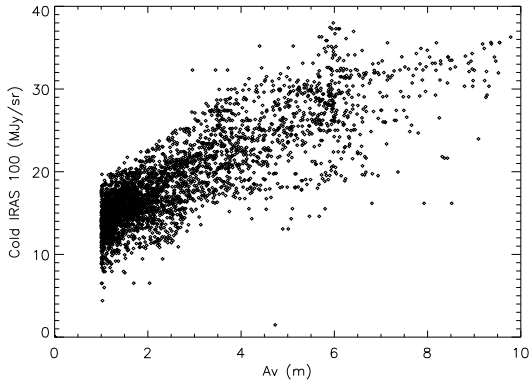,height=5.5cm}
\fi
\caption[]{Variation of cold IRAS $100\mic$ emission with visual extinction. Each point corresponds to a position on the map spaced of 1\arcmin\ in both directions}
\label{iras_av}
\end{figure}

\section{Conclusion}

The extinction map of the Chamaeleon I cloud has been significantly improved
for extinction greater than $4 \Av$ with respect to previous maps obtained from
star counts on Schmidt plates. Four distinct maxima are detected and we reach
9 visual magnitudes of extinction without degradation of the resolution. This
result has been obtained both by exploiting the massive star counts in the
$J$ band provided by DENIS, and by applying a variant of the classical star
count method which is adapted to large variations of extinction and a
wavelet analysis of the extinction map. Moreover, DENIS give us the opportunity
to investigate the cloud at a larger scale than the earlier investigations
which were limited to small regions around the visible reflection nebulae.

The comparison with the cold IRAS 100$\mic$ is striking. Each of the 3 most
important extinction maxima  corresponds to a peak of cold IRAS emission. The
IRAS flux is therefore a good indicator of extinction. We plan further
investigations of the relation between the cold IRAS emission and extinction 
when new CO observations at a comparable spatial resolution will be available.

Finally we stress the fact that our map has been derived from $J$ star counts
converted to visual extinction with $\Rv=3.1$.
This value of $3.1$ is the standard estimate for diffuse interstellar medium,
but it can reach 5.5 in dense molecular cloud cores (Whittet et al., 1987).
According to Cardelli et al. (1989) :
$$
\frac{A_{J}}{\Av}=0.4008-\frac{0.3679}{\Rv}
$$
So, we obtain 0.282 for $\Rv=3.1$ and 0.334 for $\Rv=5.5$.
This  means that we overestimate the extinction by a factor 1.18 if we choose
$\Rv=3.1$ rather than  5.5. 
Extinction is generally expressed in visual magnitudes. This choice is not
the best because the extinction law in the visible range depends on the
composition of the medium. On the other hand, the extinction law in the 
infrared seems universal. Therefore, it should be better to refer the 
extinction to near infared magnitudes.
Lastly, it appears that $J$ DENIS data with a detection
limit of $16^{\rm th}$ magnitude are better adapted to investigate the
obscuration of regions where the extinction is larger than 4 magnitudes, rather
than the Schmidt plate even at a usual detection limit of $22^{\rm nd}$ magnitudes. 

\begin{acknowledgements}
The DENIS team is warmly thanked for making this work possible and in
particular the operations team at La Silla headed by P. Fouqu\'e.
The DENIS project is supported
by the
 {\it SCIENCE  and the Human Capital and Mobility} 
plans of the European Commission under grants   CT920791 and CT940627, the
European Southern Observatory, in France by the {\it Institut National des Sciences de l'Univers}, the Education Ministery and the 
{\it Centre National de la Recherche Scientifique}, in Germany by  the State of 
Baden--Wurttemberg, 
in Spain by the DGICYT, in Italy by the Consiglio Nazionale delle Ricerche,
in Austria by the Science Fund (P8700-PHY, P10036-PHY) and Federal Ministry of
Science, Transport and the Arts, in Brazil by the Fundation for the development
of Scientific Research of the State of S\~ao Paulo (FAPESP).
\end{acknowledgements}

\end{document}